\documentclass[11pt]{article} 
\usepackage{fullpage}
\usepackage{graphicx}
\usepackage{jeffe}

\def\TRUE{\textsc{True}} 
\def\FALSE{\textsc{False}} 
\newtheorem{theorem}{Theorem} 
\newtheorem{definition}[theorem]{Definition} 
 
\begin{document} 
 
\title{Distance-2 Edge Coloring is NP-Complete} 
\date{March~1, 2002}
\author{ 
Jeff Erickson \qquad Shripad Thite \qquad David P.\@ Bunde\\ 
Department of Computer Science, University of Illinois at Urbana-Champaign\\
\{jeffe,thite,bunde\}@uiuc.edu 
} 
\maketitle 
 
\begin{abstract}
We prove that it is NP-complete to determine whether
there exists a distance-$2$ edge coloring (strong
edge coloring) with~$5$ colors of a bipartite $2$-inductive graph
with girth~$6$ and maximum degree~$3$.
\end{abstract}
 
\medskip

\noindent
Let $G$ be a simple, undirected graph.  We say that two edges 
of $G$ are within distance~$2$ of each other if either they are 
adjacent or there is some other edge that is adjacent to both of them. 
A \emph{distance-2-edge-coloring} of $G$ is an assignment of colors to 
edges so that any two edges within distance~$2$ of each other have 
distinct colors, or equivalently, a vertex-coloring of the square of 
the line graph of $G$.  If the coloring uses only $k$ colors, it is 
called a \emph{$k$-D2-edge-coloring}, and the graph $G$ is said to be 
\emph{$k$-D2-edge-colorable}.  Any $k$-D2-edge-colorable graph is also 
$(k+1)$-D2-edge-colorable.  A distance-2-edge-coloring is also known
as a \emph{strong edge coloring}.
Mahdian~\cite{mahdian00msthesis,mahdian02complexity} proved, via a
reduction from \textsc{Graph $k$-colorability}, that it is NP-complete
to determine, for every fixed $g$, whether a bipartite graph with
girth~$g$ has a strong edge coloring with~$k$ colors for $k
\ge 4$.  We present a new proof that shows that the strong edge
coloring problem is NP-complete for bipartite $2$-inductive graphs of
maximum degree~$3$.

\begin{definition}[$c$-inductive graphs]
  A graph $G$ is said to be $c$-inductive if the vertices can be
  numbered so that at most $c$ neighbors of any vertex $v$ have higher
  numbers than $v$.
\end{definition}

\begin{theorem} 
Determining whether a bipartite $2$-inductive graph with girth~$6$ and
maximum degree~$3$ is $5$-D2-edge-colorable is NP-complete.
\end{theorem} 
 
\begin{proof} 
The problem is clearly in NP since a coloring can be verified in 
polynomial time.  To prove NP-hardness, we describe a reduction from 
\textsc{Not-All-Equal-3SAT}~\cite[Problem~LO3]{GJ}. 
 
\begin{quote} 
\textsc{Instance:} A set $X = \set{x_1, x_2, \ldots, x_n}$ of 
variables, and a collection $C = \set{C_1, C_2, \ldots, C_m}$ of 
boolean clauses over $X$, each with exactly three literals. 
 
\textsc{Question:} Is there a truth assignment for $X$ such that each 
clause in $C$ has at least one true literal and at least one false 
literal? 
\end{quote} 
 
Given an instance $(X, C)$ of \textsc{Not-All-Equal-3SAT}, we will 
reduce it to a graph $G(X,C)$ that has a $5$-D2-edge-coloring 
(hereafter, a \emph{valid coloring}) if and only if $(X, C)$ has a 
satisfying assignment.  We will label the five colors $\set{T, F, 1, 
2, 3}$, where $T$ and $F$ represent the boolean values \TRUE\ and 
\FALSE. 
 
Our reduction uses three types of gadgets, all illustrated in Figure 
\ref{Fig:gadgets}. 
 
\begin{figure} 
\footnotesize\sf\centering 
\begin{tabular}{c} 
    \begin{tabular}{cc} 
	 \includegraphics[scale=0.5]{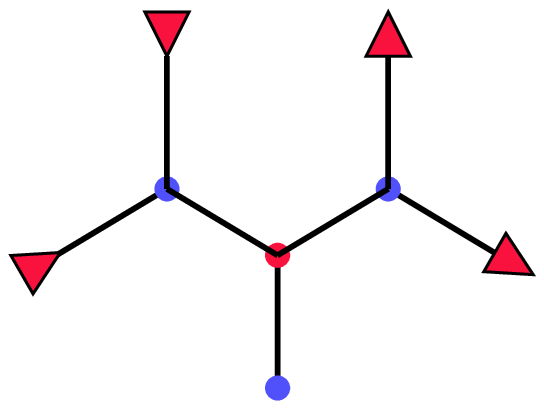} &
         \includegraphics[scale=0.5]{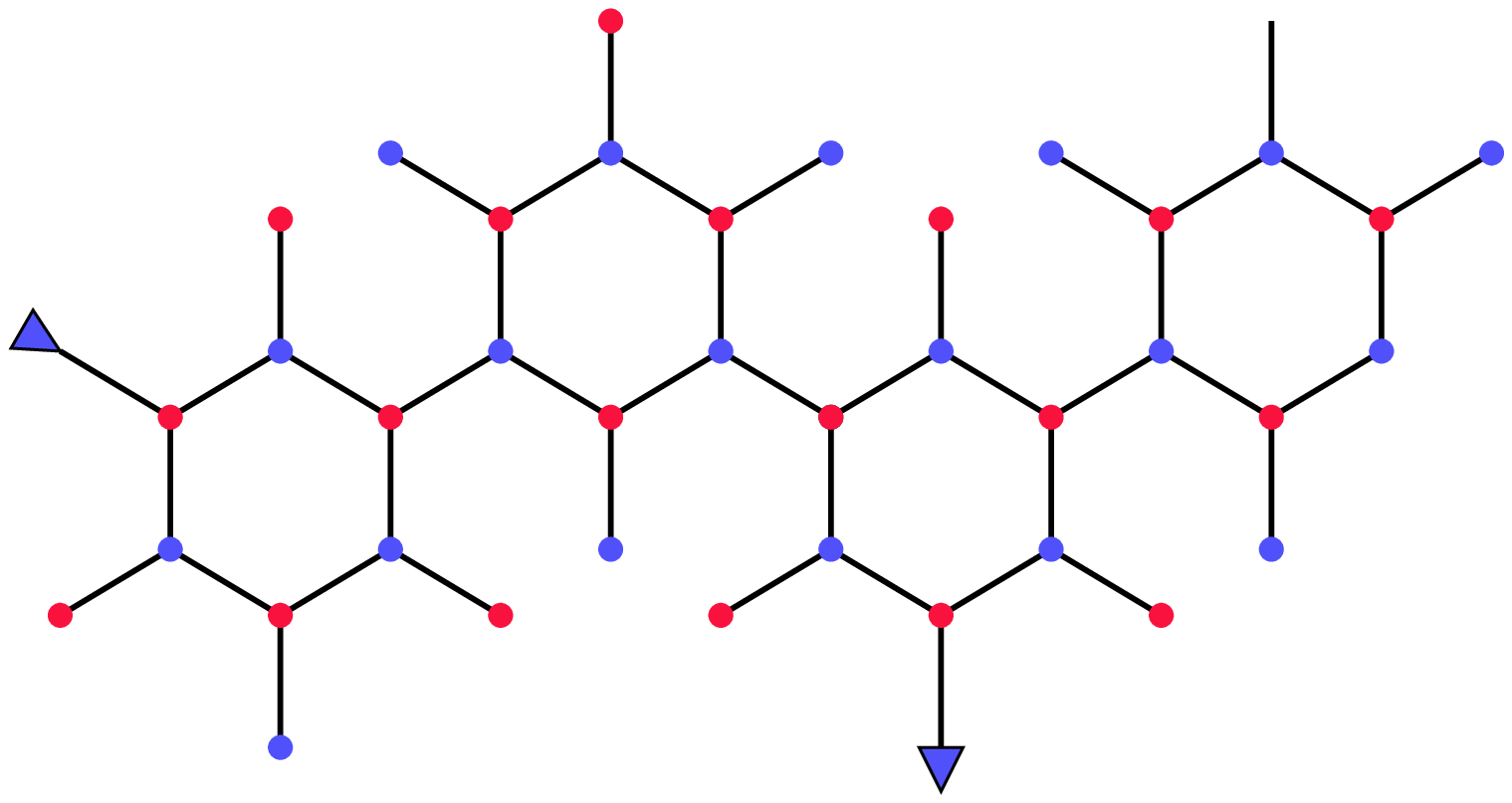} \\
	(a) & (b) 
    \end{tabular}\\ \\ 
    \includegraphics[scale=0.5]{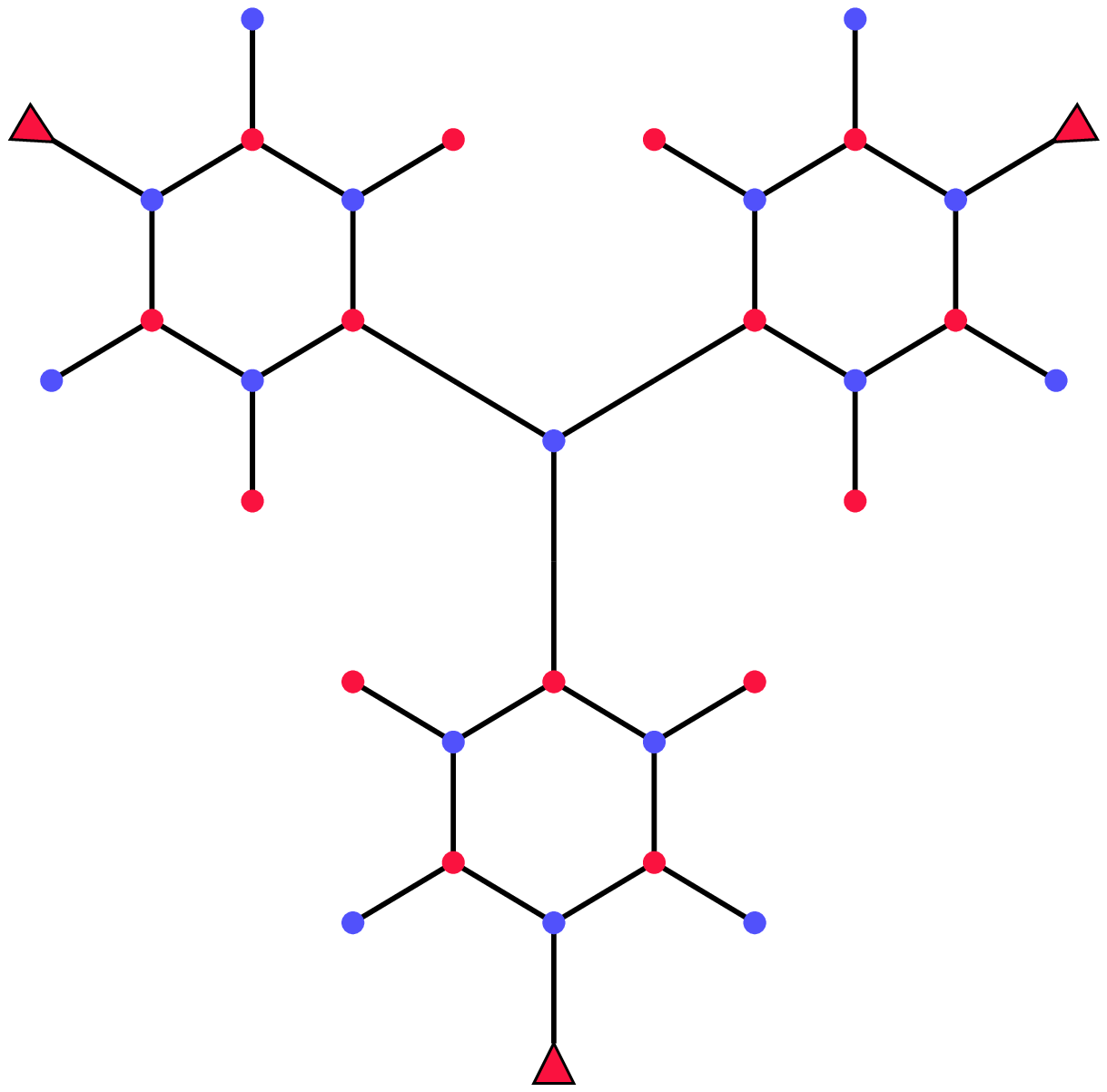} \\ 
    (c) 
\end{tabular} 
\caption{(a)  A variable gadget.  (b) A fanout gadget.  (c) A clause
gadget.  Triangles indicate input and output edges.}
\label{Fig:gadgets} 
\end{figure} 
 
\begin{itemize} 
\item 
\textbf{Fanout gadget:} This extendible gadget has a valid coloring 
with five colors such that the output edges have the same color as the 
input edges, regardless of which two colors are used on the two edges 
adjacent to the output edge.  In fact, all marked edges in the figure 
are required to be the same color.  Our reduction uses $2n+2$ fanout 
gadgets: a \emph{truth} gadget, a \emph{falsehood} gadget, and two 
\emph{literal} gadgets for each variable in $X$.  Without loss of 
generality, the output edges of the truth gadget are colored~$T$, and 
the output edges of the falsehood gadget are colored~$F$. 
 
\item 
\textbf{Variable gadget:} 
This gadget has two input edges, three internal edges, and two output 
edges.  One input edge is connected to the truth gadget, the other 
connected to the falsehood gadget.  (Thus, the truth and falsehood 
gadgets must output different colors.)  The output edges are colored 
$T$ and $F$ in any valid coloring, but either assignment of colors is 
possible.  Each output is connected to one of the literal fanout 
gadgets.  The reduction uses $n$ variable gadgets, one for each 
variable in $X$. 
 
\item 
\textbf{Clause gadget:} This gadget has three input edges, each 
connected to an output edge of the appropriate literal gadget.  The 
clause gadget has a valid coloring if and only if the three input 
edges are \emph{not} all the same color.  The reduction uses $m$ 
clause gadgets, one for each clause in~$C$. 
\end{itemize} 

Overall, our graph $G(X,C)$ has complexity $O(n+m)$ vertices and
edges, and we can easily construct it in linear time.  The graph is
bipartite because its vertices can be consistently colored with two
colors, red and blue, as shown in Figure \ref{Fig:gadgets}.  The graph
has girth~$6$ and maximum degree~$3$.  It is easy to see that the
graph is also $2$-inductive---repeatedly delete a vertex of smallest
degree, which has degree~$1$ or~$2$, and consider the vertices in the
reverse order. Any satisfying assignment for $(X,C)$ can be
transformed into a valid coloring for $G(X,C)$, and vice versa, in
$O(n+m)$ time.
\end{proof}


\end{document}